\pdfoutput=1

\documentclass[hyper]{JHEP3}

\usepackage{amsmath}
\usepackage{amssymb}
\usepackage{graphicx}

\newcommand{\Eq}[1]{(\ref{#1})} 
\newcommand{\be}{\begin{equation}}
\newcommand{\ee}{\end{equation}}
\newcommand{\bea}{\begin{eqnarray}}
\newcommand{\eea}{\end{eqnarray}}

\newcommand{\bbibitem}[1]{\bibitem{#1}} 
\newcommand{\mc}{\mathcal}
\newcommand{\mbb}{\mathbb}

\newcommand{\color}[2][]{}

\title{Non-Gaussianity in String Cosmology: A Case Study}

\author{Per Berglund$^\spadesuit$$^\heartsuit$
and Guoqin Ren$^\spadesuit$\\
$\spadesuit$ Department of Physics, University of New Hampshire, 
Durham, NH 03824, USA\\
 $\heartsuit$ PH-TH Division, CERN, CH-1211 Geneva 23, Switzerland\\
E-mail:  \email{per.berglund@unh.edu},  \email{grv2@unh.edu}}

\abstract{ 
We study non-gaussianity effects, using the $\delta N$ formalism,  in a multi-field inflationary model consisting of K\"ahler moduli derived from type IIB string compactification in the large volume limit. The analytical work in this paper mostly follows the separable potential method developed by Vernizzi and Wands.
The numerical analysis is then used in computing non-gaussianity beyond slow-roll regime. The possibility of the curvaton scenario is also discussed. We give the condition for the existence of the curvaton and calculate the non-guassianity generated by the curvaton decay in the large volume limit.}

\preprint{CERN-PH-TH/2010-235\\
UNH-10-03}

\begin{document}
\section{Introduction}
Since it was first proposed in the early 1980s, 
inflation has provided an important insight into the understanding of the very early universe~\cite{guth1981,Linde:1981mu,Albrecht:1982wi}. It also affords the possibility of probing the fundamental theories that provide a microscopic explanation of inflation such as string theory. (For a review, see~\cite{lythriotto1999, tasi}.)
In recent years, there have been some promising developments of models derived or inspired by string theory, such as type II flux compactifications~\cite{GKP,KKLT}, where scalar fields associated to the shape and size of the internal space, or to the positions of branes, serve as candidate inflaton fields~\cite{Kachru:2003sx}.
There now exists many inflationary models derived from string compactifications which correctly can account for the observational results from WMAP, such as the spectral index, $n_s$, and the ratio of tensor to scalar perturbations, see e.g. for a  review~\cite{Baumann:2009ni}.

Due to the nature of cosmological models arising from string compactifications, these inflationary models have several interesting features: (i) in general, the scalar potential is a highly non-trivial function that depends on many scalar fields and (ii) the fields are not canonically normalized, i.e., the metric is in general neither diagonal nor field independent.
These features of string theory lead to a highly coupled dynamical situation, in which, in principle, the motion of any one field impacts the evolution of the other fields.
Therefore, we need to understand how each modulus and its perturbations evolve during and after inflation, including both light and heavy fields. In this paper, we extend our earlier study~\cite{0912.1397} to include the non-gaussian fluctuations in a multi-field inflationary model arising from string theory.

Gaussian fluctuations are described by the two-point function and the corresponding power spectrum. The observed primordial perturbation to date is gaussian to a good accuracy in agreement with the predictions of inflation. However, it is possible that future experiments may observe non-gaussianity in primordial perturbation. It is therefore important to study the non-linear effects in inflationary models that can can give rise to such non-gaussian fluctuations.

In single field inflationary case, the non-linear parameters $f_{NL}$ and $g_{NL}$, which characterize the size of non-gaussianity, can be calculated in terms of the slow-roll parameters~\cite{0611075}\cite{0610210}. The result is generally small, of the order of slow-roll parameters. In  multi-field models, there are usually both heavy fields and light fields. The light fields are believed to drive inflation and heavy fields are  frozen during inflation.  It is often assumed that the dominate contribution to non-gaussianity comes from the inflaton. However, for an arbitrary scalar potential, it is not clear that whether the other (non-inflaton) fields have any sizable contribution to non-gaussinity. We would like to address this issue, at least for the model studied in this paper. 

In the inflationary scenario in which the primordial curvature perturbation originates from the inflaton, other light fields only play a role in assisting with stabilizing the potential. In the curvaton scenario, however, the curvaton (light, non-inflaton field) may have significant contribution to the primordial perturbation if its energy density grows large enough at a later time after inflation but before it decays into radiation. We explore the possibility of a curvaton scenario and compute the amount of non-gaussianity generated by the curvaton.

We focus on  a string inspired  model based on the large volume scenario~\cite{vb-pb-jc-fq}\cite{CQS}\cite{0509012}. We study different configurations of the model with different numbers of scalar fields (moduli) and with various values of the volume of the  Calabi-Yau to provide hints of the microscopic physics by connecting non-gaussianity (if observable) with the model parameters (moduli, the volume, etc). 

The outline of the paper is as follows. We first review the non-gaussian perturbations and the $\delta N$ formalism in Section~2. In Section~3, we introduce the scalar potential arising in the large volume scenario of type IIB string compactifications. We then apply the separable potential method to the above multi-field inflationary model in Section~4. In Section~5, a numerical analysis is carried out that extends the previous analytical study beyond slow-roll. Comparing the two methods, we find a good agreement in the regions where they overlap. Finally, in Section~6 we study under what conditions a curvaton may exist after the end of inflation in this type of model derived from string theory, and calculate the contributions to $f_{NL}$.

\section{Non-gaussian perturbations and the $\delta N$ formalism}

In the following, we briefly review some general facts about non-gaussian perturbations and the $\delta N$ formalism which is a powerful tool to calculate non-gaussian effects. (For more detailed discussions, see eg~\cite{0611075}\cite{0705.4096}.)

Let us define the e-folds time  
\begin{equation}
 N = \int^{t_c}_{t_*} H dt 
\end{equation}
where $t_*$ is usually chosen to be some time during inflation (the initial flat slice) and  $t_c$ is some epoch later with constant curvature perturbation (the final slice of uniform density).
In multi-field theory, the primordial curvature perturbaiton reads~\cite{0705.4096}\cite{0908.4269}\cite{1001.5259}
\begin{equation}
 \zeta(\mathbf{x})= \delta N = N_i \delta\phi^i + \frac{1}{2} N_{ij} \delta\phi^i\delta\phi^j + \frac{1}{6} N_{i j k}  \delta\phi^i\delta\phi^j\delta\phi^k + ...
\end{equation}
where $N_i, N_{ij},... $ are derivatives of the e-folds with respect to the fields $\phi^i$. $\delta\phi^i$ are evaluated on the initial (flat) slice while the derivatives of $N$ are evaluated on the unperturbed trajectory with respect to the unperturbed fields at Hubble crossing\cite{9507001}. In the later sections, we will use $_*$ to denote Hubble crossing. 

The three-point correlation function is given in terms of the bispectrum $B_\zeta(k_1,k_2,k_3)$
\begin{equation}
 \left<\zeta_{\mathbf{k_1}}\zeta_{\mathbf{k_2}}\zeta_{\mathbf{k_3}} \right> = (2\pi)^3 \delta^3(\mathbf{k_1}+\mathbf{k_2} + \mathbf{k_3})B_\zeta(k_1,k_2,k_3)
\end{equation}
where $\zeta_{\mathbf{k_i}}$ are the Fourier coefficients of  $\zeta(\mathbf{x})$. If slow-roll is satisfied, the bispectrum is completely specified by the non-linear parameter $f_{NL}(k_1,k_2,k_3)$~\cite{0802.4138}\cite{0005036}: 
\begin{equation}
 B_\zeta(k_1,k_2,k_3) =  \frac{6}{5}  f_{NL}(k_1,k_2,k_3)[P_\zeta(k_1)P_\zeta(k_2) + \text{cyclic permutations}]
\end{equation}
where $P_\zeta(k_i)$ is the power spectrum. 
Using the $\delta N$ formulation, the non-linear parameter $f_{NL}$ is given by~\cite{0506056}\cite{0504045}\cite{v-w-06}
\begin{equation}
\label{fnlmomdep}
\frac{6}{5} f_{NL}(k_1,k_2,k_3)  = \frac{k_1^3k_2^3k_3^3}{k_1^3+ k_2^3 + k_3^3}\frac{B_\zeta(k_1,k_2,k_3)}{4\pi^4 P_\zeta^2} = \frac{\sum_{i,j} N_iN_j N_{ji}}{\sum_l N_l^2} + O(r)  
\end{equation}
where $r$ is the tensor to scalar ratio. The correction $O(r)$ is a $k_i$-dependent geometric term~\cite{v-w-06}\cite{0210603}. In standard slow-roll inflation, 
\begin{equation}
 r \simeq 16 \epsilon, \quad \epsilon \sim \frac{1}{2}\frac{V_i V^i}{V^2} \ll 1
\end{equation}
Thus $O(r)$  is much less than unity due to slow-roll condition and observation constraints on $r$~\cite{1001.4635}\cite{WMAP7}. From now on we  focus on the first term in \Eq{fnlmomdep} and redefine the momentum-independent non-linear parameter 
\begin{equation}
\label{fnlred}
 f_{NL} =  \frac{5}{6} \frac{\sum_{i,j} N_iN_j N_{ji}}{(\sum_l N_l^2)^2}
\end{equation}
The four-point function has  the form 
\begin{equation}
\left<\zeta_{\mathbf{k_1}}\zeta_{\mathbf{k_2}}\zeta_{\mathbf{k_3}} \zeta_{\mathbf{k_4}}\right> = (2\pi)^3 \delta^3(\mathbf{k_1}+\mathbf{k_2} + \mathbf{k_3})T_\zeta(\mathbf{k_1},\mathbf{k_2},\mathbf{k_3}, \mathbf{k_4}) 
\end{equation}
Neglecting corrections of the order of the slow-roll parameters, the trispectrum $T_\zeta$ reads~\cite{0611075}\cite{0802.4138}\cite{1004.0818}
\begin{align*} 
T_\zeta(\mathbf{k_1},\mathbf{k_2},\mathbf{k_3}, \mathbf{k_4}) = &\tau_{NL} [P_\zeta(k_1)P_\zeta(k_2)P_\zeta(k_{14}) + 11 \text{ permutations}] \\
 &+ \frac{54}{25} g_{NL} [P_\zeta(k_2)P_\zeta(k_3)P_\zeta(k_{4}) + 3\text{ permutations}]
\end{align*}
where $k_{14} = |\mathbf{k}_1 - \mathbf{k}_4|$. Note that, unlike the bispectrum, the trispectrum depends on the directions of $k_i$'s. The parameters,  when  corrections of slow-roll order are neglected, are given by~\cite{1001.5259} 
\begin{equation}
\label{taunl}
  \tau_{NL} = \frac{N_{ij}N^{ik}N^jN_k}{(N_l N^l)^3}
 \end{equation}
\begin{equation}
\label{gnl}
  g_{NL} = \frac{25}{54} \frac{N_{ijk}N^{i}N^jN^k}{(N_l N^l)^3}
 \end{equation}
Applying the Cauchy-Schwarz inequality to \Eq{fnlred} and \Eq{taunl}, we get the following relation~\cite{0709.2545}
 \begin{equation}
  \label{c-s-ineq}
\tau_{NL} \geq \frac{36}{25}f_{NL}^2
 \end{equation}

\section{The Scalar Potential and K\"ahler Moduli Stabilization}
\label{model}

In what follows we focus on a particularly inflationary model derived from string theory consisting of multiple K\"ahler moduli, in the large volume limit (also known as the Large Volume Scenario)~\cite{vb-pb-jc-fq}\cite{CQS}.  

Supergravity in a four dimensional theory with ${\cal N}=1$ supersymmetry is completely specified by the real K\"ahler potential $\mc{K}(\varphi^i,\bar\varphi^i)$ and the holomorphic superpotential $W(\varphi^i)$. 
Focusing on the  dynamics of the scalar fields relevant for inflation, the supergravity action is (we will work in the Einstein frame, and in units where $M_P^2=1$)
\begin{equation}
S_{{\cal N}=1} = \int d^4 x \sqrt{-g} \left[ \frac{1}{2} \,R - {\cal G}_{i\bar{j}} \partial_\mu \varphi^{i} \partial^\mu \bar{\varphi}^j
- V(\varphi_i, \bar{\varphi}_i) \right]\,.
\label{eq:action}
\end{equation}
The scalar potential depends on the superpotential $W$, the K\"ahler potential $K$ as well as the K\"ahler metric ${\cal G}_{i\bar{j}} $,
\begin{eqnarray}
V(\phi_i, \bar{\phi}_i)  &=& e^{\mathcal{K}} \left({\cal G}^{i \bar{j}} D_i {W} D_{\bar{j}} \bar{{W}} -3 {W}
\bar{{W}} \right) + 
V_{\text{uplift}} \label{eq:potential}
\\
D_i {W} &=& \partial_i {W} + {W} \partial_i \mathcal{K} \\
{\cal G}_{i\bar{j}} &=& \partial_i \partial_{\bar j} \mathcal{K}
\end{eqnarray}
The derivatives $\partial_i$ and $\partial_{\bar{i}}$ differentiate with respect to the $\varphi_i$ and $\bar{\varphi}_{\bar{i}}$ dependence, respectively.   
By expanding the complex fields in terms of their real and imaginary part, we can relate the supergravity action above, (\ref{eq:action}), to the action, (\ref{action}). 
The term $V_{\text{uplift}}$  will  include the effects of supersymmetry breaking arising from other sectors of the theory.

We will demonstrate our methods in the context of Type IIB string theory compactified to four dimensions on a Calabi-Yau orientifold because the scalar potential in this case is well-understood and realistic four-dimensional models can be constructed~\cite{GKP,KKLT,vb-pb-jc-fq,CQS,DouglasDenef,Allenach}.  
After including the leading perturbative and non-perturbative corrections of string theory, the
K\"ahler potential and superpotential are given by
\begin{eqnarray}
\mathcal{K} & = & - 2 \ln \left(\mathcal{V} + \frac{ \xi \, g_s^{\frac{3}{2}}}{2 e^{\frac{3 \phi}{2}}} \right)
- \ln(-i(\tau - \bar{\tau})) - \ln \left(-i \int_{M} \Omega \wedge \bar{\Omega}\right), \nonumber \\
{W}  & = & \frac{g_s^{\frac{3}{2}} }{\sqrt{4 \pi}} \left(\frac{1}{l_s^2} \int_{M}
G_3 \wedge \Omega + \sum A_i e^{-a_i T_i} \right)
\label{eq:potentials}
\end{eqnarray}
Here $g_s$ is the string coupling, $l_s$ is the string length,  $\Omega$ is the holomorphic three-form on the Calabi-Yau manifold $M$,  $G_3$ is the background field (flux) that is chosen to thread 3-cycles in $M$ and
\be
\xi = -\frac{\zeta(3) \, \chi(M)}{2 (2 \pi)^3}
\label{xidef}
\ee
where $\chi$ is Euler number of $M$.  The axion-dilaton field is $\tau=C_0+ i \, e^{-\phi}$, and the integrals involving $\Omega$ are implicitly functions of the complex structure moduli.  The fields  $T_i = \tau_i + i b_i$ are the complexified K\"ahler moduli where $\tau_i$ is a  4-cycle volume (of the divisor $D_i\in H_4(M,\mbb{Z})$)  and $b_i$ is its axionic partner arising ultimately from the 4-form field.    Here $a_i = 2\pi/N_i$ for some integer $N_i$, for each field, that is determined by the dynamical origin of the exponentials in the superpotential ($N_i = 1$ for brane instanton contributions, $N_i > 1$ for gaugino condensates). Finally, $\mathcal{V}$ is the dimensionless classical volume of the compactification manifold $M$ (in Einstein frame, but measured in units of the string length).   In terms of the K\"ahler class $J=\sum_i t^i D_i$ (by Poincar\'{e} duality $D_i\in H^2(M,\mbb{Z}$)), with the $t^i$ measuring the areas of 2-cycles, $C_i$,
\begin{equation}
{\mathcal V} =  \int_M J^3 = \frac{1}{6} \kappa_{ijk} t^i t^j t^k~,
\label{CYvolume}
\end{equation}
where $\kappa_{ijk}$ are the intersection numbers of the manifold.   ${\mathcal V}$ should be understood as an implicit function of the complexified 4-cycle moduli $T_k$ via the relation
\begin{equation}
\tau_i = \partial_{t_i} {\mathcal V} = \frac{1}{2} \kappa_{ijk} t^j t^k~.
\label{4to2cycles}
\end{equation}

There are additional perturbative corrections to $\mathcal{K}$ in (\ref{eq:potentials}), but we have kept the terms that give the leading contributions to the scalar potential in the large $\mathcal{V}$ limit of interest to us~\cite{bergetal}.  In particular, expanding $\mathcal{K}$ to linear order in $\xi$ gives a consistent approximation in inverse powers of $\mc{V}$.   We have also assumed that all of the K\"ahler moduli $T_i$ appear in the superpotential (see~\cite{DouglasDenef} for examples) and that we use a basis of 4-cycles such that the exponential terms in ${W}$ take the form $exp(- a_i \, T_i)$.   As these exponentials arise from an instanton expansion, in order to only keep the first term as we have done, the 4-cycle volumes must be sufficiently large to ensure that $a_i T_i \gg 1$. 

Finally, the form of the term $V_{\text{uplift}}$ in (\ref{eq:potential}) depends on the kind of supersymmetry breaking effects that arise from other sectors of the theory.  
We take
\begin{equation}
V_{\text{uplift}} = \frac{\gamma}{\mc{V}^2}\,
\label{uplift}
\end{equation}
which will describe the energy of a space-filling antibrane~\cite{KKLT}, fluxes of gauge fields living on D7-branes~\cite{BKQ}, or the F-term due to a non-supersymmetric solution for the complex structure/axion-dilaton moduli~\cite{Saltman:2004sn}.

It was shown in~\cite{KKLT} that a generic choice of background fields $G_3$ causes all the complex structure moduli and the axion-dilaton to acquire string scale masses without breaking supersymmetry.  They are then decoupled from the low-energy theory and their contributions to $\mc{K}$ and ${W}$ are constants for our purposes\footnote{In the case of the F-term breaking due to the complex structure/axion-dilaton moduli~\cite{Saltman:2004sn}, the contribution of the complex structure and axion-dilaton moduli to the scalar potential does depend on the volume \eqref{uplift}.}:
\begin{eqnarray}
\mc{K} & = & - 2 \, \ln \left(\mc{V} + \frac{ \xi}{2 } \right)
-\ln \left( \frac{2}{g_s}\right) + \mc{K}_0 , \nonumber \\
{W}  & = & \frac{g_s^{\frac{3}{2}} }{\sqrt{4 \pi} } \left( W_0 + \sum_i A_i e^{-a_i T_i} \right)\,,
\label{eq:potentials2}
\end{eqnarray}
where ${\cal K}_0$ ($W_0$) is the complex structure K\"ahler potential (superpotential), evaluated at the locations where the complex structure moduli have been fixed.
It was shown in~\cite{vb-pb-jc-fq} that, when the Euler number, $\chi < 0$,  for generic values of $W_0$ (and hence of the background fluxes $G_3$), the scalar potential for the K\"ahler moduli has a minimum where the volume ${\mc V}$  of the Calabi-Yau manifold $M$ is very large -- the associated energy scale is a few orders of magnitude lower than the GUT scale.     Furthermore, in these Large Volume Scenarios  there is a natural hierarchy -- one of the K\"ahler moduli is much larger than the others and dominates the volume of the manifold,
 \begin{equation}
\tau_1 \gg \tau_2, \tau_{3} , \tau_4 \cdots
\label{modulihierarchy}
\end{equation}
which we will use to simplify the effective potential. 
For our purposes these models are also attractive because the scalar potential admits an expansion in inverse powers of the large volume $\mc{V}$.  This will allow us to carry out analytical calculations of inflation arising from K\"ahler moduli rolling towards the large volume minimum of the potential.

For transparency of the equations, we will assume  that the intersection numbers $k_{ijk}$ are such that in the basis of 4-cycles, $\tau_i$, the volume takes the diagonal form~\cite{CQS} 
\begin{equation}
\mathcal{V} = \alpha ({\tau_1}^{\frac{3}{2}}-\sum_{i=2} \lambda_i {\tau_i}^{\frac{3}{2}}) = - \alpha\sum_{i=1} \lambda_i {\tau_i}^{\frac{3}{2}}
\end{equation}
where  $\lambda_1 = -1$, and $\lambda_i, i \geq 2$ are usually positive. 

With the volume taking the above form we can explicitly compute the metric on the moduli space, 
${\cal G}_{i \bar{j}} = \partial_i \partial_{\bar{j}} \mathcal{K}$, which is needed both for the 
kinetic energy ${1\over 2} h^{ij} \dot \tau_i \dot \tau_j$, where $h_{ij}=2 {\cal G}_{i \bar{j}}$,  and for the scalar potential, $V$, appearing in the four dimensional action,
\begin{equation}\label{action}
S = \int d^4 x \sqrt{-g}[ \frac{R}{2} - \frac{1}{2}h^{ij}\partial_\mu\tau_i\partial^\mu\tau_j -V(\tau_k)]
\end{equation}
By expanding
in inverse powers of ${\cal V}$, keeping terms to $O({\cal V}^{-2})$, we obtain
\begin{equation}\label{metric}
h_{i j}= {1\over 2}\left(\frac{3\alpha \lambda_i}{8 (\mathcal{V} + \frac{\xi}{2}) {\tau_i}^{\frac{1}{2}}} \delta^{ij} + \frac{ 9 \alpha^2 \lambda_i \lambda_j \sqrt{ \tau_i \tau_j} }{8 (\mathcal{V} + \frac{\xi}{2})^2}\right)\,.
\end{equation}
With the axions minimized in the potential, 
the effective potential then becomes~\cite{0509012}
\begin{equation}\label{effpotential}
V= \sum^n_{i=2}\frac{8(a_i A_i)^2\sqrt{\tau_i}}{3\mathcal{V}\lambda_i \alpha} e^{-2 a_i \tau_i} -  \sum^n_{i=2}\frac{4 a_i A_i W_0 \tau_i}{{\mathcal{V}}^2} e^{- a_i \tau_i} + \frac{3\xi {W_0}^2}{4 {\mathcal{V}}^3} + \frac{ \gamma {W_0}^2}{{\mathcal{V}}^2}\,,
\end{equation}
where we have assumed that ${\cal K}_0$ can be chosen such that the overall scale of the potential is simplified, i.e., overall factors of $g_s$ and $2\pi$ are not present.
Here  we have expanded $V$ to $O({\cal V}^{-3})$ to include the leading $\alpha'$-corrections, $\frac{3\xi {W_0}^2}{4 {\mathcal{V}}^3}$, as well as the uplift term, $\frac{ \gamma}{{\mathcal{V}}^2}$. The parameters in the potential can be chosen and tuned under certain constraints~\cite{roulette,ourprevious,jimmy,08092982}.

To canonically normalize the metric, we can apply the following field transformations~\cite{0912.1397} 
\begin{align}
\phi^1 &= \sqrt{\frac{3\lambda_1(1 + 3\lambda_1)}{4}}\, \log(\tau_1) \label{fieldtran1}\\
\phi^i  &= \sqrt{ \frac{4 \lambda_i}{3 {\tau_1}^{\frac{3}{2}} }}{\tau_i}^{\frac{3}{4}}, \text{ } i\geq 2\label{fieldtran2}
\end{align}
by keeping terms to leading order in the expansion of inverse powers of the volume, and using that in the large volume scenario
\be
{\cal V}  \approx  \alpha {\tau_1}^{\frac{3}{2}}~.
\label{volume}
\ee
This field redefinition results in a good approximation to the canonical metric. 
After canonical normalization, the  equations of motion read
\begin{equation}
 \label{background0}
\ddot{\phi^i}+ 3 H \dot{\phi^i} + \frac{\partial V}{\partial \phi^i} = 0, \quad i =1, ..., n.
\end{equation}
To get successful inflation, we take the following steps (for more detail, see~\cite{0912.1397}). First, find the global minimum of the potential by 
\begin{equation}\label{globalminimum}
\frac{\partial V}{\partial \tau_i} = 0, \quad i=1, ..., n.
\end{equation}
Written explicitly, 
\begin{equation}
\label{1con}
 \frac{\partial V}{\partial \mathcal{V}} =0
\end{equation}

 \begin{equation}
\label{2con}
  \mathcal{V}= \frac{3\alpha \lambda_i W}{4a_iA_i}\frac{1-a_i\tau_i}{\frac{1}{4}-a_i\tau_i}\sqrt{\tau_i}e^{a_i\tau_i}, \quad i=2, ..., n.
 \end{equation}
 where \Eq{1con} is obtained by
\begin{equation}
 0 = \frac{\partial V}{\partial \tau_1}= \frac{\partial V}{\partial \mathcal{V}}\frac{\partial \mathcal{V}}{\partial \tau_1},
\end{equation}
and we apply \Eq{volume} 
\begin{equation} 
\frac{\partial \mathcal{V}}{\partial \tau_i} = 0, \quad i \geq 2 
\end{equation} 
to get \Eq{2con}. 
To have a small but positive cosmological constant, we also require that 
\begin{equation} 
V_{\text{min}} >0
\end{equation} 
In practice we want $a_i\tau_i \gg 1$ so that all higher order non-perturbative corrections of the form $e^{-m a_i \tau_i}$, with integer $m>1$, in the scalar potential are negligible and the effective potential~\Eq{effpotential} becomes a valid approximation. We find that the global minimum of the potential only exist with the parameters lying in certain regions of the parameter space. Let the global minimum be 
\begin{equation} 
V_{\text{min}}, {\tau_i}_{\text{min}}.
\end{equation} 
We have 
\begin{equation}
\label{vmina}
V_{\text{min}}= P\frac{W_0^2}{\mathcal{V}^3}
\end{equation}
where 
\begin{equation}
\label{pdef}
 P = \frac{-3}{2}\sum_{i=2}^n \frac{\alpha\lambda_i}{a_i^\frac{3}{2}} \left(\text{ln}\mathcal{V} - \text{ln} C_i\right)^\frac{3}{2} + \frac{3\xi}{4}+ \gamma \mathcal{V}
\end{equation}
and 
\begin{equation}
 C_i = \frac{3\alpha\lambda_i W_0}{4 a_i^{\frac{3}{2}} A_i}
\end{equation}

Then, for inflation to start, we displace the fields away from the global minimum along the flat direction of the potential,  and find the corresponding local minimum. Denote the values of the fields at the local minimum by 
\begin{equation} 
V_{\text{ini}}, {\tau_i}_{\text{ini}}.
\end{equation}  
which are the initial conditions of the model. 
By solving the equation of motion~\Eq{background0}, we should get successful inflation in which the fields evolve slowly toward the global minimum (${\tau_i}_{\text{min}}$). 

\section{The Inflaton Scenario}

\subsection{Separable Potential Method}
\label{spm}
In order to use the $\delta N$-formalism, we need to calculate the derivatives of the number of e-folds, $N$, with respect to the fields.  For an arbitrarily shaped potential, this can be done using numerical method which will be discussed in the next section, while the analytic treatment of non-gaussianity is known for being difficult. If the potential satisfies certain criteria~(\cite{v-w-06},\cite{0906.0767}, for example), the e-folds can be obtained by analytical integration . In the model introduced in Section \ref{model}, it has been shown~\cite{0906.3711}\cite{0912.1397} that the non-inflaton (or heavy) fields are frozen before the end of inflation and only the light fields evolve during inflation. Thus, before the end of inflation, we can ignore the dependence of the potential on the heavy fields and treat the potential as a function of the light fields only. 

Let us assume that we have a volume modulus, $\tau_1$, a heavy modulus, $\tau_2$, and two light moduli, $\tau_3$ and $\tau_4$ (as in~\cite{0912.1397}), corresponding to the  canonically normalized fields $\phi_3$ and $\phi_4$, \Eq{fieldtran1}-\Eq{fieldtran2}. Furthermore, if we suppose that $\tau_1$ and $\tau_2$ are frozen during inflation, the potential \Eq{effpotential} can be seperated as two functions each depending only on one of the light fields ($\phi_3, \phi_4$):
\begin{equation}
V=  U(\phi_3) + W(\phi_4)
\end{equation}
The terms which contain the frozen fields have been absorbed into $U(\phi_3)$ and $ W(\phi_4)$. 

Next we will follow the \emph{separable potential method} developed by Vernizzi and Wands~\cite{v-w-06} to calculate the derivatives of the e-folds.  In the canonical frame, assuming the background fields only have time dependence, the background equations of motion read
\begin{equation}
 \label{background}
\ddot{\phi^a}+ 3 H \dot{\phi^a} + V^{,a} = 0, \quad a = 3, 4.
\end{equation}
where 
\begin{equation}
 V^{,a} = \frac{\partial V}{\partial \phi^a} = \begin{cases}
                                                 U', \quad a = 3;\\
                                                W', \quad a = 4.
                                               \end{cases}
\end{equation}
And the Einstein field equations are\footnote{Again we set the planck mass $M^2_{pl} =1$.}
\begin{equation}
 3H^2 = \frac{1}{2} g^{\mu\nu} \partial_\mu\phi^a\partial_\nu\phi_a +V
\end{equation}
\begin{equation}
 \dot{H} = -\frac{1}{2} g^{\mu\nu} \partial_\mu\phi^a\partial_\nu\phi_a
\end{equation}
Using the slow-roll approximation, we have
\begin{align}
\label{slow-roll}
 \quad H^2 &\simeq \frac{V}{3} \\
 \frac{1}{H} &\simeq - \frac{3 \dot{\phi}_3}{U'} \simeq  - \frac{3 \dot{\phi}_4}{W'}
 \label{identity}
\end{align}
By integrating~(\ref{identity}) we then get 
\begin{equation}
 \int \frac{d \phi_3}{U'} = \int \frac{d \phi_4}{W'} + C 
\end{equation}
The number of e-folds becomes
\begin{align}
 N & = \int^c_* \frac{H^2}{H} dt \nonumber\\
 & = \int^c_* \frac{U}{3H} dt + \int^c_* \frac{W}{3H} dt  \nonumber\\
 & = - \int^c_* \frac{U}{U'} d\phi_3 - \int^c_* \frac{W}{W'} d\phi_4
\end{align}
Here $_*$ and $_c$ denote the time at Hubble exit and some time after Hubble exit, respectively. We usually choose the latter as the time, $t_c$, for some constant Hubble parameter (uniform energy density hypersuface), $H_c$. Then we  let $t_*$ vary and compute the derivative of $N$ with respect to the initial fields at $t_*$. 
The results, derived by Vernizizi and David~\cite{v-w-06}, are, with $N_{i*}=\partial N/\partial \phi^i|_{t=t_*}$,  
\begin{align}\label{N3}
 N_{3*} &= \frac{1}{\sqrt{2\epsilon_{3*}}}\frac{U_*+ Z_c}{V_*},\\ \label{N4}
 N_{4*} &= \frac{1}{\sqrt{2\epsilon_{4*}}}\frac{W_*- Z_c}{V_*},\\ \label{N33}
 N_{33*} &= 1 - \frac{\eta_{3*}}{2\epsilon_{3*}}\frac{U_*+ Z_c}{V_*} + \frac{1}{V_*\sqrt{2\epsilon_{3*}}}\frac{\partial Z_c}{\partial \phi_{3*}},\\ \label{N44}
N_{44*} &= 1 - \frac{\eta_{4*}}{2\epsilon_{4*}}\frac{W_*- Z_c}{V_*} - \frac{1}{V_*\sqrt{2\epsilon_{4*}}}\frac{\partial Z_c}{\partial \phi_{4*}},\\ \label{N34}
N_{34*} &=   \frac{1}{V_*\sqrt{2\epsilon_{3*}}}\frac{\partial Z_c}{\partial \phi_{4*}} = - \frac{1}{V_*\sqrt{2\epsilon_{4*}}}\frac{\partial Z_c}{\partial \phi_{3*}}.
\end{align}
where 
\begin{equation} \label{slowrollZ}
 \epsilon_{a} = \frac{1}{2}\left(\frac{V_{,a}}{V}\right)^2, \quad
\eta_{a} = \frac{V_{,aa}}{V} ,\quad
 Z = \frac{W\epsilon_{3} - U\epsilon_{4}}{\epsilon_{3} + \epsilon_{4}}
\end{equation}
and 
\begin{equation}\label{Zc}
 \sqrt{\epsilon_{3*}}\frac{\partial Z_c}{\partial \phi_{3*}} = - \sqrt{\epsilon_{4*}}\frac{\partial Z_c}{\partial \phi_{4*}} = -\sqrt{2} V_*\frac{{V_c}^2}{{V_*}^2}\frac{\epsilon_{3c}\epsilon_{4c}}{\epsilon_{3c} + \epsilon_{4c}}\left(1 - \frac{\epsilon_{4c}\eta_{3c} + \epsilon_{3c}\eta_{4c}}{(\epsilon_{3c} + \epsilon_{4c})^2}\right)
\end{equation}

Note that this method relies on the slow-roll approximation~\Eq{slow-roll}. It requires that the final slice at $t_c$ must lie within the slow-roll regime. To calculate the amount of non-gaussianity generated near and after the end of inflation, one needs to find an alternative method valid beyond slow-roll. For example, the authors of~\cite{0906.0767} proposed an analytic method, valid for certain classes of inflation models with separable Hubble functions, which can be used to study non-gaussianity after inflation ends. Although their analysis applies to certain types of potentials with exponential terms, the detailed conditions are not satisfied for the scalar potential~(\ref{effpotential}). In section~5, we present a numerical analysis valid beyond the slow-roll regime.

\subsection{Estimate of $f_{NL}$} 
\label{estfnl}
\paragraph{Two light fields case:}
In the model discussed in Section \ref{model}, the two light fields ($\phi_3, \phi_4$) serve as candidates for inflaton. We know make a rough estimate of $f_{NL}$ in this case. 
Let us assume that $\phi^4$ is the assisting field and thus lighter than the inflaton $\phi^3$.  During slow-roll we would then expect that  
\begin{equation}
 \epsilon_{4}< \epsilon_{3} \ll 1 ,| \eta_{4}| < |\eta_{3}| \ll 1, |\eta_i| \gg\epsilon_i, \quad i = 3, 4
\end{equation}
since it follows that $|V_{,ii}| \gg V_i$ from the flatness of the potential \Eq{effpotential} and the fact that both fields are light, $|\eta_i| =|\frac{V_{,ii}}{V}| \ll 1$. 
Let 
\begin{equation}
 |W(\phi^4)| = p|U(\phi^3)|, \quad p > 0,
\end{equation}
 \begin{equation}
 \epsilon_4 = s \epsilon_3,\quad 0 < s < 1,
\end{equation}
and 
\begin{equation}
 |\eta_4| = q|\eta_3|, \quad 0< q <1.
\end{equation}
Although, in general, $p, s, q$ are functions of time, they are not expected to change dramatically in the slow-roll regime. In particular, we can treat $p$ as a constant due to the flatness of the potential. Furthermore, while the actual value of $p$ depends on the specific values chosen for the parameters in $W(\phi^4)$ and $U(\phi^3)$, respectively, one finds that $p\sim O(1)$ in a typical model. Using the separable potential method introduced in the previous section, we get 
\begin{equation}
 N_{3*}= \frac{1}{\sqrt{2\epsilon_{3*}}} \frac{p}{1 + s_c}, \quad N_{4*} = \frac{s_c}{\sqrt{s_*}} N_{3*}
\end{equation}
\begin{equation}
 N_{33*} \approx 1 - \frac{p}{1 + s_c} \frac{\eta_{3*}}{2 \epsilon_{3*}} +  \frac{s_c(s_c + q_c)}{(1+s_c)^3}\frac{\eta_{3c}}{\epsilon_{3*}}
\end{equation}
\begin{equation}
 N_{44*} \approx 1 - \frac{s_c q_*}{s_*}\frac{p}{1 + s_c} \frac{\eta_{3*}}{2 \epsilon_{3*}} + \frac{s_c }{s_*} \frac{s_c(s_c + q_c)}{(1+s_c)^3}\frac{\eta_{3c}}{\epsilon_{3*}}
\end{equation}
\begin{equation}
 N_{34*} \approx - \frac{1}{\sqrt{s_*}}\frac{s_c(s_c + q_c)}{(1+s_c)^3}\frac{\eta_{3c}}{\epsilon_{3*}} 
\end{equation}
The non-linear parameter $f_{NL}$, defined by \Eq{fnlred} and evaluated at $t_*$ for a fixed $t_c$, then becomes
\begin{equation}
 f_{NL} \approx x \eta_{3*} + y \epsilon_{3*} 
\end{equation}
where the coefficients are 
\begin{align*}
 x &= \frac{1}{2}(1+\frac{s_c^2}{s_*})^{-2}\left[\frac{s_c(s_c + q_c)}{p^2(1+s_c)}(1- \frac{s_c}{s_*})^2 \frac{\eta_{3c}}{\eta_{3*}} - \frac{1 + s_c}{p}(1 + q_*\frac{s_c^3 }{s_*^2}) \right], \\
 y & = \frac{2}{p^2}\frac{(1 + s_c)^2}{1+\frac{s_c^2}{s_*}}
\end{align*}
both of which are of $ O(\frac{1}{p^2})$ because of the slowly changing $s$ and $q$.

\paragraph{Mixed case:}
Consider a model in which there is one heavy fields, $\phi_2$, and one light field $\phi_3$ (inflaton) in addition to the volume field, $\phi_1$. As usual, $\phi_1$ is frozen during inflation.  But we drop the constraint that the heavy field $\phi_2$ has been stabilized. We want to see how much the heavy field contributes to the non-gaussianity. The potential \Eq{effpotential}  can be separated as 
\begin{equation} 
 V = \tilde{U}(\phi_2) + \tilde{W}(\phi_3) 
 \end{equation} 
where 
\begin{equation} 
\tilde{U}(\phi_2)=\frac{8(a_2 A_2)^2\sqrt{\tau_2}}{3\mathcal{V}\lambda_2 \alpha} e^{-2 a_2 \tau_2} -  \frac{4 a_2 A_2 W_0 \tau_2}{{\mathcal{V}}^2} e^{- a_2 \tau_2} 
\end{equation} 
\begin{equation} 
\tilde{W}(\phi_3)=\frac{8(a_3 A_3)^2\sqrt{\tau_3}}{3\mathcal{V}\lambda_3 \alpha} e^{-2 a_3 \tau_3} -  \frac{4 a_3 A_3 W_0 \tau_3}{{\mathcal{V}}^2} e^{- a_3 \tau_3}
\end{equation} 
First, since $\phi_2$ is heavy and $\phi_3$ is light, 
\begin{equation}
\label{assumption2}
\eta_2 = \frac{V_{,22}}{V} \sim O(1), \eta_3 = \frac{V_{,33}}{V} \ll 1; \quad \epsilon_2 \ll \epsilon_3
\end{equation}
Because the inflaton  $\phi_3$ is usually displaced far away along the flat direction of the potential, we would expect that 
\begin{equation} 
 \tilde{W}(\phi_3) \ll \tilde{U}(\phi_2) 
 \end{equation} 
In addition, we assume that during slow-roll
\begin{equation}
 \epsilon_c \sim \epsilon_*, \quad \eta_c \sim \eta_* 
\end{equation}
We then have from~(\ref{N3}) and~(\ref{N4})
\begin{equation}
 N_{2*} \sim \frac{\sqrt{\epsilon_{2*}}}{\epsilon_{3*}}, \quad N_{3*} \sim \frac{1}{\sqrt{\epsilon_{3*}}}
\end{equation} 
Since  $\phi_2$ is heavy and $\phi_3$ is light it follows from~(\ref{assumption2}) that
\begin{equation} 
N_{3*} \gg N_{2*}
\end{equation} 
and~(\ref{Zc}) can be written
\begin{equation} 
\frac{\partial Z_c}{\partial \phi_{2*}} \sim  V \frac{\epsilon_{2c}}{\epsilon_{3c}}\frac{\eta_{2c}}{\sqrt{\epsilon_{2*}}}, \quad \frac{\partial Z_c}{\partial \phi_{3*}} \sim  V \frac{\epsilon_{2c}}{\epsilon_{3c}}\frac{\eta_{2c}}{\sqrt{\epsilon_{3*}}}
\end{equation} 
Using these results we can estimate the expressions for $N_{ij}$,  and (\ref{N33}-\ref{N34}) becomes
\begin{equation} N_{22*} \sim O(1), \quad N_{33} \sim -\frac{\eta_{3*}}{\epsilon_{3*}}, \quad  N_{23*} \sim \frac{\epsilon_{2c}}{\epsilon_{3c}}\frac{\eta_{2c}}{\sqrt{\epsilon_{2*}\epsilon_{3*}}}
\end{equation} 
from which it follows that
\begin{equation} 
N_{33*} \sim \frac{\eta_{3*}}{\eta_{2c}} \sqrt{\frac{\epsilon_{3*}}{\epsilon_{2c}} }  N_{23*}
\label{Nijrelation}
\end{equation} 
By assumption, \Eq{assumption2}, the slow-roll factors satisfy 
\begin{equation} 
\frac{\epsilon_{3c}}{\epsilon_{2c}}\gg 1\,,\quad \sqrt{\frac{\epsilon_{3*}}{\epsilon_{2*}} }\gg 1\,,\quad \frac{\eta_{3c}}{\epsilon_{2c}} \gg 1, \quad  \frac{1}{\eta_{2c}} \sim O(1),
\label{slowrollhierarchy}
\end{equation} 
Using~(\ref{slowrollhierarchy}) in~(\ref{Nijrelation}) we then arrive at the following hierarchy among the $N_{ij*}$
\begin{equation}
N_{33*} \gg N_{23*} \gg N_{22*}
\end{equation}
Therefore, 
\begin{equation} 
f_{NL} = \frac{5}{6} \frac{\sum_{i,j =2}^3 N_iN_j N_{ji}}{(\sum_l N_l^2)^2} \approx  \frac{5}{6} \frac{  N_{33}}{ N_3^2} \sim  O (\eta_3) \ll 1
\end{equation}
with the dominant contribution coming from the light field $\phi_3$.

In summary, both examples discussed in this section yield $ f_{NL} \sim O(\eta)\ll 1$ (where $\eta$ is the slow-roll parameters for the inflaton). Since $|\eta|>>\epsilon$ for the inflaton, the dominant contribution to $f_{NL}$ is indeed the momentum independent term in~(\ref{fnlmomdep}). The result is similar to that of the standard slow-roll inflation, see for example~\cite{0611075}.

\subsection{Explicit Setups}
\label{spmr}
We now perform an explicit calculation of $f_{NL}$ given by~(\ref{fnlred}). To evaluate  $N_i$ and $N_{ij}$ in~(\ref{N3}-\ref{Zc}), we need to compute 
$Z$ at fixed $t_c$. This is done by solving the equations of motion for the background fields numerically. Below we present to examples corresponding to two different values for the volume of the Calabi-Yau.
\paragraph{Example 1}

Let us choose the parameters in the potential~\Eq{effpotential} as 
\[  \alpha=\frac{1}{9\sqrt{2}}, a_2=\frac{2\pi}{300}, a_3=\frac{2\pi}{100}, a_4=\frac{2\pi}{100} A_2=0.2, A_3=0.001, A_4=0.001. \]
\[\lambda_1 = -1,  \lambda_2=0.1 , \lambda_3=0.005, \lambda_4=0.005, W=500, \xi=40, \gamma=9.75 \times 10^{-6}.  \]
Choose the initial conditions to be 
\[\tau_{1}(0)=76212.1, \tau_{2}(0)=246.99, \tau_{3}(0)= 472.42, \tau_{3}(0)= 491.54,\]
\[\dot{\tau_{1}}(0) =\dot{\tau_{2}}(0) = 0, \dot{\tau_{3}}(0) =-1.72\times10^{-19}, \dot{\tau_{4}}(0) =-1.5\times10^{-19}.\]
The volume in this setup is $\mathcal{V} \sim 10^6$ which is within a reasonable range $10^3 - 10^8$~\cite{1005.4840}. This will give 60 e-folds before the end of inflation. The nonlinear coefficients are~\footnote{The different values of $N$ are computed at correspondingly different, constant, values of the Hubble parameter, $H_c$.}
\begin{figure}[ht]
\begin{center}
\begin{tabular}{c|c|c|c|c|c|c}
 $N(H_c)$& $N$=20 & $N$=30 & $N$=40 & $N$=50 & $N$=55 & $N$=59 \\
\hline\hline \\
$f_{NL}$  & 0.0146 & 0.0147 & 0.0147 & 0.0147 & 0.0147 & 0.0147 \\
\hline \\
$\tau_{NL}$       & 0.000308 & 0.000312& 0.000312& 0.000311 & 0.000311 & 0.000311 \\
\end{tabular}
\end{center}
\label{table:spm1}
\end{figure}

\paragraph{Example 2} 

As a second example, we choose the parameters such that the volume is relatively small, $\mathcal{V} \sim 10^3$.
\[  \alpha=\frac{1}{9\sqrt{2}}, a_2=\frac{2\pi}{80}, a_3=\frac{2\pi}{80},  A_2=0.04, A_3=1.2\times10^{-4}, A_4=1.2\times10^{-4}. \]
\[\lambda_1 = -1,  \lambda_2=1 , \lambda_3=0.01, \lambda_4=0.01, W=1, \xi=35, \gamma=2.65 \times 10^{-3}.  \]

\[\tau_{1}(0)=1781.356, \tau_{2}(0)=51.039, \tau_{3}(0)= 282, \tau_{3}(0)= 285,\]
\[\dot{\tau_{1}}(0) =\dot{\tau_{2}}(0) = 0, \dot{\tau_{3}}(0) =-1.40948\times10^{-9}, \dot{\tau_{4}}(0) =-1.21344\times10^{-9}.\]
The inflation lasts for $N= 62.5$ e-folds. 
\begin{figure}[ht]
\begin{center}
\begin{tabular}{c|c|c|c|c|c|c}
  $N(H_c)$  & $N$=20 & $N$=30 & $N$=40 & $N$=50 & $N$=55 & $N$=59 \\
\hline\hline \\
$f_{NL}$  & 0.0171 & 0.0167& 0.0148& 0.0108 & 0.0121 & 0.0170\\
\hline \\
$\tau_{NL}$       & 0.000423 & 0.000429& 0.000465& 0.000464 & 0.000333 & 0.000423 \\
\end{tabular}
\end{center}
\label{table:spm2}
\end{figure}

One can check that the above explicit results are consistent with the conclusion in section (4.2), that $f_{NL} \sim O(\eta)$.

\section{Numerical Methods}
\label{nm}
\subsection{The Finite Difference Method}

Numerically it is straightforward to solve the equations of motion for the background fields without applying slow-roll approximation~\cite{0912.1397}.  The advantage of the numerical method is that we do not need to rely on slow-roll approximation (although we still need to assume slow-roll at Hubble exit~\cite{v-w-06}) and no assumption  about the shape of the potential is needed. 

We will use the \emph{finite difference method}~\cite{leveque} to calculate the derivatives of $N= N(\phi_1, ..., \phi_n;H_c)$\footnote{ Here $\phi_i$ are understood to be the field values at the Hubble exit.} up to the second order beyond the slow-roll regime. 

\paragraph{First Order derivative}

The finite difference method gives 
\begin{equation}
 N_i  = \frac{1}{2h_i}[N(\phi_1, ...,\phi_i+h_i,... \phi_n) -N(\phi_1, ...,\phi_i-h_i,... \phi_n)] + O(h^2)
\end{equation}

\paragraph{Second Order derivative}

When $ i = j$,
\begin{equation}
 N_{ii} = \frac{1}{h_i^2}[N(\phi_1, ...,\phi_i+h_i,... \phi_n) -2N(\phi_1, ..., \phi_n) + N(\phi_1, ...,\phi_i-h_i,... \phi_n)] + O(h^2)
\end{equation}
and when $i \neq j$,
\begin{align}
 N_{ij} &= \frac{1}{4h_ih_j}[N(\phi_1, ...,\phi_i+h_i,...,\phi_j+h_j,... \phi_n) - N(\phi_1, ...,\phi_i+h_i,...,\phi_j-h_j,... \phi_n) \nonumber \\
&-N(\phi_1, ...,\phi_i-h_i,...,\phi_j+h_j,... \phi_n)+ N(\phi_1, ...,\phi_i-h_i,...,\phi_j-h_j,... \phi_n) ] + O(h^2)
\end{align}
Once the  $N_i$'s and $N_{ij}$'s are obtained, we are ready to calculate the non-gaussianity using the $\delta N$-formalism discussed previously\footnote{We do not compute the third order derivative of $N$, and thus $g_{NL}$,  since the term containing it is proportional to $O(h^3)$ which is very small and the error bars can be relatively large.}.

\subsection{Example}
\label{nr}
We numerically solve the background equations of motion for the model introduced in Section \ref{model}. Then we use the $\delta N$ formalism  to calculate the non-linear parameters $f_{NL}$ and $\tau_{NL}$. 
The parameters in the potential~\Eq{effpotential} and initial conditions are chosen the same as in Section \ref{spmr}, Example 1. 

\begin{figure}[ht]
\begin{center}
\begin{tabular}{c|c|c|c|c|c|c|c}
 $N(H_c)$ & $N$=20 & $N$=30 & $N$=40 & $N$=50 & $N$=55 & $N$=59 & $N$=60.4  \\
\hline\hline \\
$f_{NL}$  & 0.00874 & 0.0125 & 0.0142 & 0.0143 & 0.0143 & 0.0143 &0.0143  \\
\hline \\
$\tau_{NL}$    &   0.000300 & 0.000274& 0.000292& 0.000295 & 0.000296 & 0.000362 & 0.000346 \\
\end{tabular}
\end{center}
\label{table:nu}
\end{figure}

These are very close to the results obtained by the analytical method in Section \ref{spmr}. Remarkably,  notice that the values of the nonlinear parameters does not change much near ($N=59$) and after ($N=60.4$) the end of inflation when slow-roll condition breaks down. It is reasonable to suspect that the non-gaussianity evolves very slowly through inflation and even preheating era. In practice, we may just use the separable potential method  to compute non-gaussianity under slow-roll condition and use the result as an approximation to those in regimes beyond slow-roll. 

\section{Curvaton Scenario}
So far we have only considered the inflationary scenario in which we assume that the non-gaussianity is generated by the inflaton. However, it is necessary to investigate the possibility of a curvaton scenario which does not affect the dynamics during inflation but may play a major role in the oscillation stage. 
\subsection{Curvaton evolution}

In a multi-field inflationary model, there will in general be several light fields, with one of them, called the inflaton, 
$\phi$, dominating the dynamics of inflation. Other light fields, on the other hand, have very little effect during inflation and are usually neglected. However, under certain circumstances, a light field other than the inflaton may be identified as the curvaton, $\sigma$, which sometimes generates significant non-gaussianity after the end of inflation~\cite{plb524}\cite{0206026}.

After the end of inflation, the inflaton quickly starts to oscillate about its potential minimum. It then decays into radiation (photon) when its decay rate $\Gamma_{\phi} \simeq H$, where the decay rate $\Gamma_{\phi}$ can be calculated once the coupled Lagrangian is given. During the oscillation process, if $\Gamma_{\phi}> \Gamma_\sigma$, the inflaton will decay first, leaving the curvaton as the only light field\footnote{We assume that there are no other light fields, such as those associated with cold dark matter, etc.}. Right after the inflaton decays into radiation, the curvaton energy density is still subdominant. However, the massless radiation decreases faster, $\sim \frac{1}{a^4}$, than the massive particles associated with the curvaton, $\sim \frac{1}{a^3}$, as the universe expands. Thus the relative energy density of the  non-relativistic curvaton may increase until it decays into radiation, at which point it may even dominate the total energy density. 

\subsection{The existence of the curvaton}
We now turn to address the question whether the curvaton scenario can occue in the class of models constructed in section~3.
For simplicity, we assume that during inflation all the fields, except $\phi_1$ and $\phi_n$, stay close to their VEVs and are thus heavy. We can write the potential as

\begin{equation}
\label{Vre}
 V \sim V_0 + V_1 + V_n
\end{equation}
 where $V_1$ and $V_n$ are potentials for $\phi_1$ and $\phi_n$, and $V_0$ is the (almost) constant part of the potential due to the (almost) frozen $\phi_i$, with $1<i<n$.

If the curvaton exists, its mass must be less than the Hubble parameter. Thus the quadratic potential $V_1 $ should be small compare to $V_0$.  The inflaton $\phi_n$ is displaced far away from its VEV, and its potential is suppressed by orders of $\frac{1}{\mathcal{V}}$  
\begin{equation}
 V_n = \frac{8(a_n A_n)^2\sqrt{\tau_n}}{3\mathcal{V}\lambda_n \alpha} e^{-2 a_n \tau_n} -  \frac{4 a_n A_n W_0 \tau_n}{{\mathcal{V}}^2} e^{- a_n \tau_n} \sim \frac{1}{\mathcal{V}^{3+\beta}}, \quad \beta>0, 
\end{equation}
 negligible if compared to $V_0$. Here $\beta$ depends on how far away $\phi_n$ has been displaced from its minimum.

 As a result, \Eq{Vre} is dominated by $V_0$
\begin{equation}
\label{v0}
 V \simeq V_0 = P_0\frac{W_0^2}{\mathcal{V}^3}
\end{equation}
where 
\begin{align}
P_0 &= -\frac{3}{2}\sum_{i=2}^{n-1} \alpha \lambda_i\left<\tau_i\right>^\frac{3}{2} + \frac{3}{4}\xi + \gamma\mathcal{V} \label{p0}\\
 \left<\tau_i\right> & \simeq a_i^{-1} \left(\text{ln}\mathcal{V}-\text{ln}C_i\right)\,,\quad  C_i = \frac{3\alpha\lambda_i W_0}{4 a_i^{\frac{3}{2}} A_i}
\end{align}
where $\left<\tau_i\right>$ is the value of the $i$th moduli at its minimum and the uplifting parameter $\gamma \mathcal{V} \sim O(1)$, see Section \ref{model}. 

Near the potential minimum, the masses of the canonicalized fields, $\phi_1$ and $\phi_i, i\geq 2$, given by \Eq{fieldtran1} and \Eq{fieldtran2}  are 
\begin{align}
\label{mass1}
m_1^2 &= Q_1  \frac{W_0^2}{\mathcal{V}^3}, \\
\label{massi}
m_i^2 &= Q_i\frac{W_0^2}{\mathcal{V}^2}, \quad i\geq 2.
\end{align}
where the coefficients $Q_i$ are given by
\begin{align}
 Q_1 &= - \frac{63}{4}\sum_{i=2}^n  \alpha\lambda_i  \left<\tau_i\right> ^{\frac{3}{2}} + \frac{81}{8} \xi + 6\gamma \mathcal{V} \\
 Q_i &= -\frac{5}{4} + 4 a_i^{-1}\left<\tau_i\right>  + 4 a_i^{-2}\left<\tau_i\right> ^2
\end{align}
Note that the mass of the inflaton, $\phi_n$, after inflation is $O({\cal V}^{-2})$ while during inflation the mass goes like $O({\cal V}^{-2-\beta})$

As expected, the  fields $\phi_i, 2\leq i\leq n-1$ are heavier than the Hubble parameter in the large volume limit
\begin{equation}
m_i^2 \sim \frac{1}{\mathcal{V}^2} > H^2 \simeq \frac{1}{3}V_0 \sim  \frac{1}{\mathcal{V}^3}
\end{equation} 
by \Eq{vmina}. 

Since the field $\phi_1$ is our candidate for the curvaton field, it should be lighter than the Hubble parameter, i.e.,  
\begin{equation}
\label{conlight}
 0< Q_1 <  \frac{1}{3}P_0,
\end{equation}
More explicitly,
\begin{align}
\frac{61}{4}\sum_{i=2}^n  \alpha\lambda_i  \left<\tau_i\right> ^{\frac{3}{2}} & + \frac{1}{2}\alpha\lambda_n \left<\tau_n\right> ^{\frac{3}{2}} > \frac{79}{8} \xi + \frac{17}{3}\gamma \mathcal{V} \label{req1}\\
 \frac{63}{4}\sum_{i=2}^n  \alpha\lambda_i  \left<\tau_i\right> ^{\frac{3}{2}} &< \frac{81}{8} \xi + 6\gamma \mathcal{V}\label{req2}
\end{align}
From~(\ref{req1})  and~(\ref{req2}) it then follows that
\begin{equation}
\alpha\lambda_n \left<\tau_n\right> ^{\frac{3}{2}}  > \frac{1}{7}\xi - \frac{20}{9} \gamma \mathcal{V}
\end{equation}

In the simpliest setup where all the fields $\phi_i, i\geq 2$ are identical in parameters($\lambda_i, a_i$, etc), we get from \Eq{req1} and \Eq{req2}
\begin{equation}
 n < 1.55
\end{equation}
which is not possible since the integer $n$ has to be at least 2. Hence a different setup than the simpliest one is needed to satisfy the curvaton condition. 

Equations~\Eq{req1} and \Eq{req2} are the \emph{necessary} conditions for the existence of the curvaton scenario.  They are very restrictive, however, as shown by the simple example above, and some fine tuning is required to  satisfy \Eq{req1} and \Eq{req2}.

\subsection{The decay rate}
It has been shown that the there exists a Lagrangian arising from D7-branes wrapping a small four-cycle, whose complexified volume is the modulus $\tau$,
\begin{equation}
\label{interactinglag}
\mathcal{L}_g = - \frac{\lambda_{\tau} }{4 M_{Pl}}\tau F_{\mu\nu} F^{\mu\nu}
\end{equation}
where $ \lambda_{\tau}$ is the coupling of $\tau$ to the gauge field (photon)~\cite{0705.3460}. 

The other parts of the Lagrangian can obtained by quadratic expansion around the potential minimum. By canonically normalizing the kinetic terms and diagonalizing the mass matrix terms
\begin{equation}
\mathcal{L}_0= -V_\text{min}+ \frac{1}{2} \partial_\mu \psi_i \partial^\mu \psi^i -  \frac{1}{2} m_{i}^2 \psi_i\psi^i
\end{equation}
where $\psi_i$ are the canonically normalized fields and also eigenfunctions of the mass matrix. 

In what follows, we consider a model consisting of multiple moduli: $\tau_1, \tau_2, ..., \tau_n$, where $\tau_1$ is the large four-cycle and all other moduli are small. Typically, most of the small cycles are close to their vevs and thus are heavy during inflation. Only the inflaton is displaced far from its vev. Let the inflaton be $\tau_n$. So the  relevent moduli here will be $\tau_1$ and $\tau_n$ and other moduli play the role of stabilizing the potential.  Starting from the Lagrangian in \Eq{action}, it is possible to simultaneously diagonalize  the kinetic terms and the mass matrix terms under the assumption that the mass matrix is independent of the fields, which is a good approximation close to the minimum of the potential. For simplicity, let us  first diagonalize the kinetic terms using the field transformations \Eq{fieldtran1} and \Eq{fieldtran2}. The Lagrangian reads
\begin{align}
 \mathcal{L}_0 = &-V_\text{min}+ \frac{1}{2} \partial_\mu \hat{\phi_1} \partial^\mu \hat{\phi_1} + \frac{1}{2} \partial_\mu \hat{\phi_n} \partial^\mu \hat{\phi_n}-  \frac{1}{2} m_{1}^2 \hat{\phi_1}^2 -  \frac{1}{2} m_{n}^2 \hat{\phi_n}^2 -m_{1n}^2 \hat{\phi_1}\hat{\phi_n}\nonumber 
\end{align}
where 
\begin{equation}
 m_{1n}^2 = \frac{\partial^2 V}{\partial \phi_1 \partial \phi_n} = Q_{1,n} \frac{W_0^2}{\mathcal{V}^\frac{5}{2}}
\end{equation}
\begin{equation}
 Q_{1,n} \simeq -3\sqrt{2\alpha\lambda_n}a_n \left<\tau_n\right>^\frac{7}{4},
\end{equation}
and the effective fields  $\hat{\phi_i} = \phi_i -\left<\phi_i\right>$ represent the oscillation amplitude of the field about its potential minimum. 

From now on we omit the hat over $\phi$. Following~\cite{1005.5076}, we calculate the eigenvalues and eigenvectors of the mass metrix
\begin{equation}
 M^2 = \begin{pmatrix}
        m_1^2 & m_{1n}^2 \\
 m_{1n}^2  & m_n^2
\end{pmatrix}
\end{equation}

\begin{equation}
 M^2 v_i = e_i v_i, \quad i =1,2.
\end{equation}
where the eigenvectors $v_i$ are normalized such that $v_i^T v_j =\delta_{ij}$. 

The transformation takes the form
 
 \begin{equation}
 \begin{pmatrix}
 \phi_1\\
 \phi_n
 \end{pmatrix}
  = \begin{pmatrix}
        v_1 
\end{pmatrix} \psi_1
+ \begin{pmatrix}
        v_2 
\end{pmatrix} \psi_2
\end{equation}
In the large volume limit, we find that 
\begin{align}
\phi_1 &\simeq \psi_1 + O\left(\mathcal{V}^{-\frac{1}{2}}\right)\psi_2 \simeq \psi_1 \\ 
\phi_2 &\simeq O\left(\mathcal{V}^{-\frac{1}{2}}\right) \psi_1 + \psi_2 \simeq \psi_2
\end{align}
which means that we can approximate the Lagrangian by 
\begin{equation}
\mathcal{L}_0\simeq -V_\text{min}+ \frac{1}{2} \partial_\mu \phi_1 \partial^\mu \phi_1 + \frac{1}{2} \partial_\mu \phi_n \partial^\mu \phi_n-  \frac{1}{2} m_{1}^2 \phi_1^2 -  \frac{1}{2} m_{n}^2 \phi_n^2 
\end{equation}


Similarly, we can rewrite the Lagrangian for the gauge sector~\cite{0705.3460}
\begin{equation}
 \label{oriinterlag}
\mathcal{L}_g= - \frac{k }{4 M_{Pl}}\tau F_{\mu\nu} F^{\mu\nu} = - \frac{k}{4 M_{Pl}}(\hat{\tau_n} + \left<\tau_n\right>) F_{\mu\nu} F^{\mu\nu}
\end{equation}
 where $k$ is a normalization factor. We set $\tau = \tau_n$ since the D7 branes only wrap the small four-cycle $\tau_n$~\cite{1005.4840}. 
  In terms of the (approximately) canonically normalized fields $\phi_i$,  the gauge field (radiation) Lagrangian~(\ref{oriinterlag}) takes the form 
 \begin{equation}
 \mathcal{L}_g = -\frac{1}{4} F_{\mu\nu} F^{\mu\nu}  - \sum_{i=1}^n\frac{\lambda_{\phi_i} }{4 M_{Pl}}\phi_i F_{\mu\nu} F^{\mu\nu}
 \end{equation}
which corresponds to
\begin{align}
 k = &\left<\tau_n\right>^{-1}, \\
 \lambda_{\phi_1} &= \frac{\sqrt{6}}{2}, \quad \lambda_{\phi_n} = \left(\frac{3\mathcal{V}}{4\alpha \lambda_n }\right)^{\frac{1}{2}} \left<\tau_n\right>^{-\frac{3}{4}},\quad \lambda_{\phi_i}\approx 0,\quad i=2,\ldots,n-1.
\end{align}

The complete Lagrangian reads
\begin{align}
\mathcal{L}= & -V_\text{min}+ \frac{1}{2} \partial_\mu \phi_1 \partial^\mu \phi_1 + \frac{1}{2} \partial_\mu \phi_n \partial^\mu \phi_n-  \frac{1}{2} m_{1}^2 \phi_1^2 -  \frac{1}{2} m_{n}^2 \phi_n^2 \nonumber\\
 &-\frac{1}{4} F_{\mu\nu} F^{\mu\nu} - \frac{\lambda_{\phi_1} }{4 M_{Pl}}\phi_1 F_{\mu\nu} F^{\mu\nu}- \frac{\lambda_{\phi_n} }{4 M_{Pl}}\phi_n F_{\mu\nu} F^{\mu\nu}\label{lagt}
\end{align}

From the Lagrangian \Eq{lagt}, it is straightforward to get the decay rates 
\begin{equation}
\Gamma_{\phi \to \gamma\gamma} = \frac{\lambda_\phi^2 m_\phi^3}{64\pi M_{Pl}^2}
\end{equation}
i.e.,
\begin{equation}
 \label{dec1}
\Gamma_{\phi_1 \to \gamma\gamma} = \frac{3W_0^3}{128\pi M_{Pl}^2}Q_1^\frac{3}{2}\frac{1}{\mathcal{V}^\frac{9}{2}}
\end{equation}
\begin{equation}
 \label{dec2}
\Gamma_{\phi_n \to \gamma\gamma} =\frac{3 W_0^3}{256\pi M_p^2\alpha \lambda_n}\left(\frac{Q_n}{\left<\tau_n\right>}\right)^\frac{3}{2} \frac{1}{\mathcal{V}^2}
\end{equation}
Thus, 
\begin{equation}
\Gamma_1 \sim \frac{1}{\mathcal{V}^\frac{9}{2}} \ll \Gamma_n \sim \frac{1}{\mathcal{V}^2}
\end{equation}

This indicates that the curvaton $\phi_1$ decay indeed occurs some time after decay of  the inflaton. The amount of non-gaussianity generated by the curvaton is determined by its relative energy at the time of decay, which will be shown in the next section. 

\subsection{The nonlinear parameter}
The curvaton starts to decay when the Hubble parameter drops below the decay rate of the curvaton~\footnote{In what follows, the curvaton in our scenario, $\phi_1$, is relabeled $\sigma$ to conform with previous work on curvatons in the literature.}
\begin{equation}
H \sim \Gamma_{\sigma \to \gamma\gamma}
\end{equation}

Using the sudden decay approximation (assuming the decay happens instantaneously), the nonlinear parameter of the curvaton perturbation can be shown to be~\cite{0504045}\cite{0607627}
\begin{equation}
\label{nlparacurv}
f_{NL} = \frac{5}{4r_{\text{dec}}} \left(1 + \frac{g g''}{g'^2}\right) -\frac{5}{3} - \frac{5r_{\text{dec}}}{6}
\end{equation}
where the dimensionless ratio
\begin{equation}
\label{rd}
r_{\text{dec}} = \frac{3 \rho_{\sigma\text{dec}}} {3 \rho_{\sigma\text{dec}} + 4 \rho_{\gamma\text{dec}}} 
\end{equation}
and $\rho_{\sigma\text{dec}}$ and $\rho_{\gamma\text{dec}}$ are the energy density for the curvaton  and radiation at the time when the curvaton decays, respectively. 

The function $g$ characterizes the dependence of the curvaton, $\sigma (=\phi_1)$, at the beginning of its oscilation,  on its value at Hubble crossing, $\sigma_*$, i.e., $\sigma = g(\sigma_*)$ . Assuming the absence of the nonlinear evolution of the curvaton,   we have $g'' =0$ and
\begin{equation}
\label{nlparacurvsim}
f_{NL} = \frac{5}{4r_{\text{dec}}} -\frac{5}{3} - \frac{5r_{\text{dec}}}{6}
\end{equation}

If the curvaton energy density dominates the total energy density when it decays, the corresponding nonlinear parameter is
\begin{equation}
f_{NL} \sim - \frac{5}{4}
\end{equation}
On the other hand, if $r_{\text{dec}}\ll 1$, then $f_{NL} \gg 1$. Note that if $r_{\text{dec}} \approx 0.58$, $f_{NL}\approx 0$.

The initial energy density of the curvaton $\sigma$,  
when it begins to oscillate, is
\begin{equation}
 \rho_\sigma \simeq \frac{1}{2} m_\sigma^2 \sigma^2
\end{equation}
where $\sigma$ is the oscillation amplitude of the curvaton. To estimate its value, we use the arguments similar to~\cite{1005.4840}\cite{0308015}. Analogous to the Hawking radiation in black holes,  the quantum fluctuation $\delta\sigma$ of the light field $\sigma$ during inflation in de Sitter space has the power spectrum~\cite{0907.5424} 
\begin{equation}
P_{\delta\sigma} = \left<|\delta\sigma|^2\right>= \left(\frac{H_\star}{2\pi}\right)^2 = T_H^2
\end{equation}
where $T_H$ is the Hawking temperature and the label $\star$ denotes the Hubble exit; and
\begin{equation}
H_\star^2 \simeq \frac{1}{3}P_0\frac{W_0^2}{\mathcal{V}^3}
\end{equation}
by \Eq{v0}. 
It indicates that the amplitude of quantum fluctuation
\begin{equation}
\delta\sigma \sim T_H =\frac{H_\star}{2\pi}
\end{equation}

The amount of quantum fluctuation is comparable to the classical (slow roll) motion when 
\begin{equation}
\delta\sigma \sim \dot{\sigma} \delta t_\star \sim \frac{V_{,\sigma}}{H_\star} H_\star^{-1} 
\end{equation}
where the slow-roll condition of the light field $\sigma$ has been used, and $\delta t_\star = H_\star^{-1} $ is the change in time  during one e-fold. We view the onset of the quantum regime as the time when the oscillation takes place. The typical (initial) value of $\sigma$  constraint by the quantum fluctuations thus satisfies the conditon

\begin{equation}
 \frac{\partial V}{\partial \sigma} \simeq H_\star^3,
\end{equation}

Since the potential is quadratic under assumption, the value of $\sigma$ reads
\begin{equation}
\sigma \simeq \frac{V_{\star}^{\frac{3}{2}}}{m_\sigma^2}
\end{equation}
where $V_{\star} = V_0 \simeq 3H^2_\star$.

The initial ratio between the curvaton energy density and the total energy density is 
\begin{equation}
\Omega_{\text{in}} = \frac{\rho_\sigma}{\rho_{\text{tot}}}\sim \frac{\frac{1}{2}m_\sigma^2 \sigma^2}{3H^2_{\text{in}} M_{pl}^2} = \frac{V_{\star}^3}{6 m_\sigma^4}
\end{equation}
where 
\begin{equation}
 H_{\text{in}} = m_\sigma
\end{equation}
and we set $M_{pl} =1$ as usual.

Since 
\begin{equation}
\label{masssigma}
m^2_\sigma = m_1^2 = Q_1 \frac{W_0^2}{\mathcal{V}^3},
\end{equation}
we have 
\begin{equation}
 \Omega_{\text{in}}= \frac{P_0^3 W_0^2}{6 Q_1^2}\frac{1}{\mathcal{V}^3}
\end{equation}

Now assume that the oscillation stage only lasts for a few e-folds ($\Delta N = \int^\text{dec}_{\text{in}} H dt$). We should have  $\Omega_{\text{in}} \ll  e^{-\Delta N} < 1$.  Then, under the sudden decay approximation, the ratio  $r_{\text{dec}} $ of Eq.~\Eq{rd}  can be related to  $\Omega_{\text{in}}$ by~\cite{0211602}
\begin{align}
 r_{\text{dec}} &\simeq \frac{3}{4}\Omega_{\text{in}}(1-\Omega_{\text{in}})^{-\frac{3}{4}}\left(\frac{H_{\text{in}}}{\Gamma}\right)^{\frac{1}{2}} \nonumber \\
 &\simeq \frac{3}{4}\Omega_{\text{in}}\left(\frac{m_\sigma}{\Gamma}\right)^{\frac{1}{2}}\nonumber \\
 & \simeq \frac{\sqrt{6\pi}}{3}  \frac{P_0^3 W_0^{\frac{5}{2}}}{Q_1^{\frac{5}{2}}} \frac{1}{\mathcal{V}^{\frac{3}{2}}}
\end{align}

In terms of $H_\star$, 
\begin{equation}
 r_{\text{dec}} \sim W_0^{\frac{3}{2}} \frac{H_\star}{M_{pl}}
\end{equation}
where we temporarily restore the Planck mass. Notice that the Hubble parameter during inflation is generally much smaller than the Planck mass. Given the fact that $W_0$ should not be very large,  the ratio $ r_{\text{dec}}$ can be quite small. By \Eq{nlparacurvsim},  this will give rise to a large positive $f_{\text{NL}} >1$, see also~\cite{1005.4840}. 

\section{Conclusion}

In the first half of this paper, we discussed the inflaton scenario. Both the analytical method (separable potential) and the numerical method (finite difference) have been used to calculate the nonlinear parameters via the $\delta N$ formalism. These two methods agree with each other very well in the string inflation model introduced in Section \ref{model}. We conclude that, although the analytical method is only valid under slow-roll, it  can be a good estimate for regimes beyond slow-roll. The nonlinear parameters we get are very small which is typical for  slow-roll inflationary models. The results do not vary much for different volume sizes. Instead, as shown in Section \ref{estfnl}, they mostly depend on the slow-roll parameters. The amount of non-gaussinaity ($f_{NL} \sim 0.01$) are unlikely to be detected by the major experiments that will be carried out~\cite{lyth-liddle-09}\cite{0807.1770}. However, a certain type of cosmological probe has been proposed~\cite{0610257}, by which this amount of non-gassianity is reachable. 

After discussing the inflaton scenario, we present the conditions for existence of the curvaton scenario. To satisfy those very restrictive conditions, the parameters need to be fine tuned appropriately. The non-gaussianity in the curvaton scenario is given in terms of the parameters in the potential and the Calabi-Yau volume. A rough estimate shows that the non-gaussianity can be quite large for a typical setup. These large non-gaussianity effects in curvaton scenarios have also recently been computed in closely related type IIB flux compactifications~\cite{1005.4840}.

\acknowledgments

We thank C.~Byrnes and L.~Verde for useful discussions. P.B. would also like to thank the Theory Division at CERN for hospitality during this project.
This work has been carried out with support from the NSF CAREER grant PHY-0645686.
P.B. acknowledges additional support from
the University of New Hampshire through its Faculty Scholars Award
Program.

\end{document}